\documentclass[conference]{IEEEtran}
\IEEEoverridecommandlockouts
\usepackage{cite}
\usepackage{amsmath,amssymb,amsfonts}
\usepackage{algorithmic}
\usepackage{graphicx}
\usepackage{textcomp}
\usepackage{xcolor}
\usepackage{algorithmic}
\usepackage[ruled,vlined,linesnumbered]{algorithm2e}
\usepackage{listings}
\usepackage{graphicx}
\usepackage{verbatim}
\usepackage[most]{tcolorbox}
\usepackage{subcaption} 
\def\BibTeX{{\rm B\kern-.05em{\sc i\kern-.025em b}\kern-.08em
    T\kern-.1667em\lower.7ex\hbox{E}\kern-.125emX}}

\begin{document}

\title{Advancing Autonomous Incident Response: Leveraging LLMs and Cyber Threat Intelligence}

\author{\IEEEauthorblockN{Amine Tellache$^{1,2}$, Abdelaziz Amara Korba$^{2}$, Amdjed Mokhtari$^{1}$, Horea Moldovan$^{1}$, Yacine Ghamri-Doudane$^{2}$}
\IEEEauthorblockA{\textit{$^{1}$OODRIVE-Trusted Cloud Solutions, 75010 Paris, France} \\
\textit{$^{2}$L3i Lab, University of La Rochelle, 17000 La Rochelle, France.}\\
emails: \{a.tellache@oodrive.com, a.mokhtari@oodrive.com, abdelaziz.amara\_korba@univ-lr.fr, \\ h.moldovan@oodrive.com, yacine.ghamri@univ-lr.fr\}
}
}

\maketitle

\begin{abstract}
Effective incident response (IR) is critical for mitigating cyber threats, yet security teams are overwhelmed by alert fatigue, high false-positive rates, and the vast volume of unstructured Cyber Threat Intelligence (CTI) documents. While CTI holds immense potential for enriching security operations, its extensive and fragmented nature makes manual analysis time-consuming and resource-intensive. To bridge this gap, we introduce a novel Retrieval-Augmented Generation (RAG)-based framework that leverages Large Language Models (LLMs) to automate and enhance IR by integrating dynamically retrieved CTI. Our approach introduces a hybrid retrieval mechanism that combines NLP-based similarity searches within a CTI vector database with standardized queries to external CTI platforms, facilitating context-aware enrichment of security alerts. The augmented intelligence is then leveraged by an LLM-powered response generation module, which formulates precise, actionable, and contextually relevant incident mitigation strategies. We propose a dual evaluation paradigm, wherein automated assessment using an auxiliary LLM is systematically cross-validated by cybersecurity experts. Empirical validation on real-world and simulated alerts demonstrates that our approach enhances the accuracy, contextualization, and efficiency of IR, alleviating analyst workload and reducing response latency. This work underscores the potential of LLM-driven CTI fusion in advancing autonomous security operations and establishing a foundation for intelligent, adaptive cybersecurity frameworks.
\end{abstract}

\begin{IEEEkeywords}
Large Language Models (LLM),  Cyber Threat Intelligence (CTI), Incident Response (IR), Retrieval-Augmented Generation (RAG).
\end{IEEEkeywords}

\section{INTRODUCTION} \label{sec:intro}

With the ever-increasing sophistication of cyber threats, organizations struggle to respond efficiently to security incidents, leading to significant financial and operational consequences.
Incident response \cite{mughal2022building} is a structured and strategic process for identifying and handling cyberattacks, aimed at reducing damage, recovery time, and overall costs. IR involves several specialized teams: SOC analysts handle monitoring and initial triage; CTI analysts provide strategic threat intelligence to contextualize events; and finally incident responders manage and remediate the detected incidents. Despite this structure, teams face major challenges, starting with detection. SOC analysts deal with alert fatigue due to high volumes—averaging 4,484 alerts per day—and spend nearly three hours daily on manual triage \cite{noauthor_securitymagazine}. Alert enrichment with threat intelligence adds further delays and resource demands. Modern multi-cloud environments (e.g., AWS, GCP, Azure) introduce additional complexity with fragmented security models, requiring specialized skills and slowing response times. Meanwhile, generic incident response playbooks often lack actionable guidance, especially for less experienced teams, limiting their ability to learn from past incidents and respond effectively.

To bridge these gaps, organizations have turned to Cyber Threat Intelligence (CTI) \cite{sun2023cyber} as a pivotal resource for enhancing their IR capabilities.
CTI involves the collection, analysis, and dissemination of information about cyber threats and vulnerabilities, enabling organizations to understand better and respond to potential risks. Within SOC teams, CTI is primarily used to enrich and contextualize security alerts, providing crucial insights for improved prioritization and understanding. This enables teams to quickly identify the nature and source of an attack, assess potential damage, and make informed recommendations. It also provides insights into the latest threats and tactics used by attackers, which helps teams better prepare for and respond to incidents. Subsequently, IR teams rely on this enriched intelligence to craft efficient and targeted responses, underlining the importance of CTI in the overall cybersecurity ecosystem. However, without automated systems, searching, reading, and correlating CTI reports requires additional manual effort. Moreover, CTI data is vast, heterogeneous, and comes from diverse sources, making its management and analysis complex. Effectively utilizing CTI requires highly skilled personnel and advanced technological tools, which can pose challenges in terms of cost and skilled-resources availability. Furthermore, the ability to quickly analyze CTI data and take preventive or corrective actions may be limited, particularly while facing real-time attacks.


Given these challenges, there is a pressing need for advanced automation tools. Large Language Models (LLMs) have transformed multiple domains~\cite{zhao2023survey, chang2024survey, kasneci2023chatgpt} with advanced language processing, automation, and data analysis, making them invaluable in many fields. These strengths make LLMs particularly valuable in cybersecurity, where they have shown promising results in diverse applications, including dataset generation \cite{liu2024api}, threat detection \cite{liu2024api,diaf2024beyond}, response \cite{sladic2024llm}, Cyber threat intelligence \cite{fayyazi2023uses}, and even attack \cite{iyengar2023large}. 
In the CTI domain specifically, LLMs are increasingly employed to address the challenge of processing heterogeneous data from diverse sources. Multiple solutions have been proposed in this context. For instance, the Mitre ATT\&CK project \cite{noauthor_tram} launched the Threat Report ATT\&CK Mapper (TRAM), designed to simplify the analysis of CTI reports and extraction of Tactics, Techniques, and Procedures. Alves et al. \cite{alves2022leveraging} utilized different BERT model variants with optimized hyperparameters to identify the most effective model for TTP classification. Similarly, Fayyazi et al. \cite{fayyazi2023uses} compared direct use of LLMs (GPT-3.5, Bard) with supervised fine-tuning (SFT) of smaller LLMs (BERT, SecureBERT) to interpret ambiguous cyberattack descriptions, demonstrating that fine-tuned small-scale LLMs like SecureBERT outperform direct use in precise classification of cybersecurity tactics. In addition, Peng et al. \cite{peng2024ctisum} introduced CTISum, a new benchmark specifically designed for CTI summarization tasks with capabilities for attack process summarization as a subtask. Tseng et al. \cite{tseng2024using} proposed an LLM-based solution to automate threat intelligence workflows within SOCs. Their system extracts Indicators of Compromise using LLMs, applies a filtering and cleaning mechanism via a multi-LLM voting system, and generates Regex rules and relationship graphs, enhancing preventive measures. 
While most state-of-the-art approaches in CTI leverage LLMs for passive tasks such as summarization and mapping, their role in active incident response remains limited. 
We strongly believe that LLMs can extend their actual usage to serve as a powerful tool, offering exceptional capabilities in processing, analyzing, and synthesizing vast amounts of CTI data to improve incident response efficiency, automating the entire workflow from report analysis to tailored response planning. However, there are significant challenges, particularly the high costs of fine-tuning these models for specialized CTI contexts. Additionally, maintaining the ability to continuously adapt to the ever-evolving landscape of new attacks and emerging CTI can be both time-consuming and expensive.

To address these challenges, we propose a novel real-time incident response approach based on LLMs. This system enables security teams to effectively leverage Cyber Threat Intelligence to respond to incidents. This model is capable of accurately enriching events and alerts using CTI, and then rapidly producing effective actions in response to specific threats, drawing on vast and diverse CTI sources. The model's architecture will incorporate and integrate CTI without requiring costly fine-tuning for each update, through the Retrieval-Augmented Generation technique. The effectiveness of this approach has been rigorously evaluated using a novel automated method, powered by LLMs, and cross-validated by cybersecurity experts. This evaluation assesses key metrics such as answer relevance, context relevance, and groundedness across two datasets: real-world alerts and simulated alerts.

In summary, the main contributions of this paper are:
\begin{itemize}
\item A novel RAG-based incident response model that effectively leverages Cyber Threat Intelligence to enrich security alerts and generate precise, context-aware response actions. Our approach supports the continuous integration of new CTI data without requiring model fine-tuning for each update, ensuring adaptability to emerging threats.

\item We propose an innovative retrieval within our RAG system that combines two complementary search techniques to contextualize and enrich incidents: The model can performs standard searches on platforms through CTI APIs like VirusTotal \cite{noauthor_vx_nodate} to retrieve standardized contexts and correlate the incident with private databases. Subsequently, it employs NLP-based similarity searches to identify relevant documents or text segments in the CTI vector database, enabling the correlation of incidents with historical cases.


\item A rigorous evaluation framework integrating automated assessments powered by LLMs with manual expert validation on real-world and simulated SIEM (Security Information and Event Management) alerts. The simulated alerts are generated by executing real attack scenarios in controlled environments. The results, assessed across multiple dimensions such as response accuracy, contextual relevance, and data groundedness, demonstrate the effectiveness of our approach in enhancing incident response.

\end{itemize}

The remainder of this paper is outlined as follows. The proposed RAG-based Incident Response architecture is described in Section~\ref{sec:SOL}. Section~\ref{sec:exp} presents the experimental results and analysis. Finally, Section~\ref{sec:con} concludes the paper and points out future directions.

\section{Proposed Approach} \label{sec:SOL}

\begin{figure*}[hbt!]
  \centering
\begin{subfigure}[b]{\textwidth}
        \includegraphics[width=\textwidth]{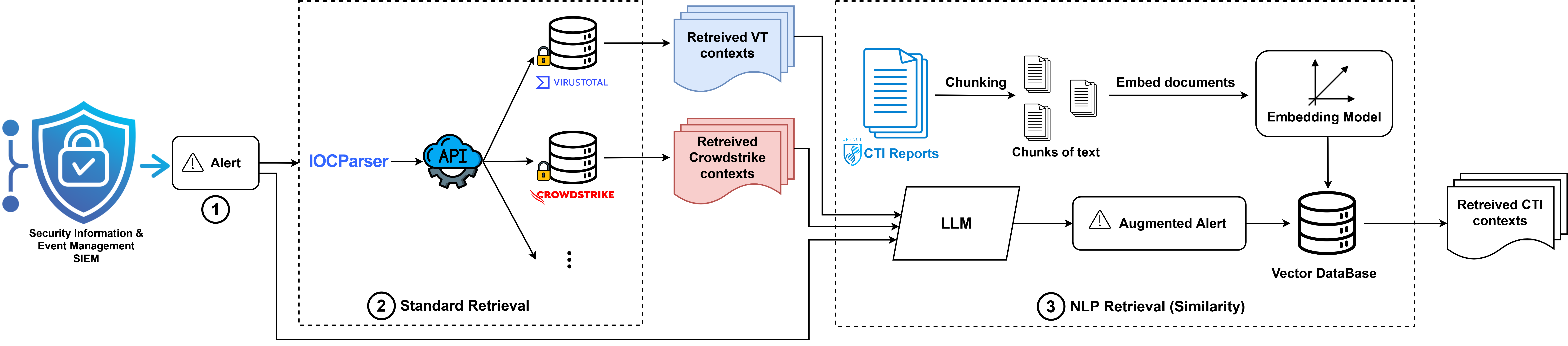}
        \caption{RAG Retrieval component}
        \label{fig:RAG-Retrieval-component}
    \end{subfigure}
    
    \vspace{0.0cm} 
    
    \begin{subfigure}[b]{\textwidth}
        \includegraphics[width=0.78\textwidth]{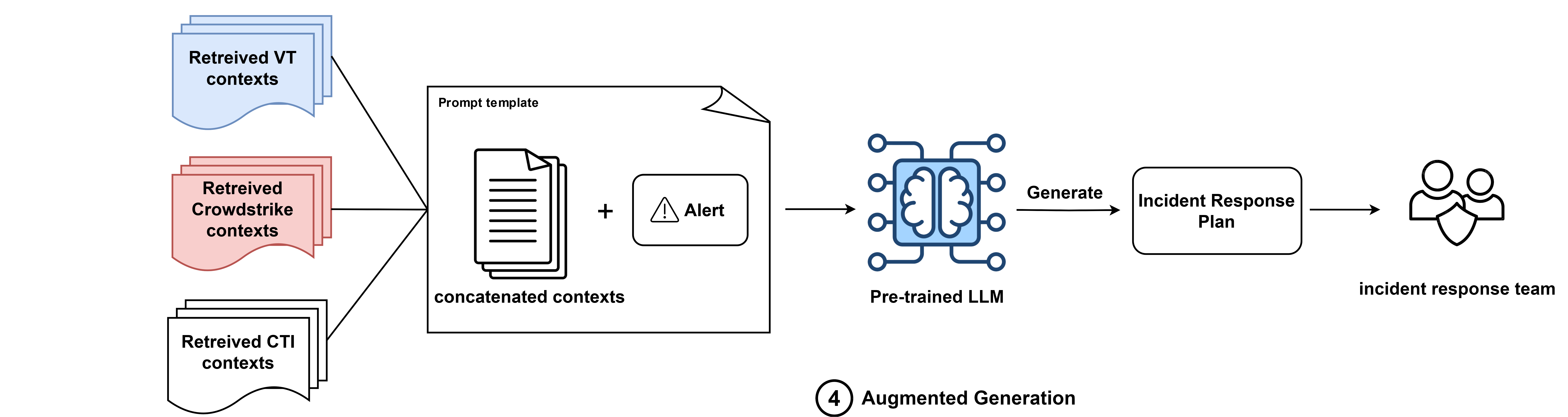}
        \caption{RAG Augmented generation component}
        \label{fig:RAG-Augmented-generation-component}
    \end{subfigure}
    
  \caption{Proposed Retrieval-Augmented Generation RAG Incident Response Architecture}
  \label{fig:Proposed-RAG-Architecture}
\end{figure*}


\subsection{Overview of the proposed system}

We propose a RAG-based incident response architecture (Figure~\ref{fig:Proposed-RAG-Architecture}) designed to automate and streamline the workflows of SOC analysts, CTI analysts, and incident responders. The system takes as input a security incident description, with a focus on SIEM alerts (Figure~\ref{fig:RAG-Retrieval-component}, Part 1), and operates in two main phases: retrieval and augmented generation. In the Retrieval phase, the model searches for relevant threat intelligence to enrich the alert. Given the inherent access restrictions in CTI, we introduce a tailored retrieval method optimized for this context. The model first performs structured searches via CTI platform APIs, such as VirusTotal, to extract contextual information from private databases. It then employs NLP-based similarity techniques to identify relevant documents or text segments within CTI vector databases, enabling correlation between current incidents and historical cases. In the augmented generation phase, the language model synthesizes accurate and actionable responses using the retrieved data. Our approach continuously integrates new CTI data by embedding reports into a vector database and using additional private datasets, eliminating the need for fine-tuning. This strategy significantly reduces both computational overhead and operational costs, ensuring adaptability to evolving threats. The following sections detail the architecture’s main components, including knowledge base construction, the retrieval mechanism, and the generation process.

\subsection{Building the Knowledge Base}
To build the Knowledge Base for the NLP-based search, we initially utilized CTI documents from publicly available collections of Advanced Persistent Threat (APT) and cybercriminal campaign reports \cite{noauthor_vx_nodate}. In the future, we plan to integrate this process directly with the OpenCTI platform \cite{noauthor_opencti_nodate} to enable seamless data sharing and integration. The process starts by loading all the CTI documents, cleaning, and transforming PDF files into text. Then, text splitters break large documents into smaller chunks, making them easier to index and search, as larger chunks can be difficult to handle in models with limited context windows. These chunks are transformed into vectors using embeddings, and then stored in a specialized database called a vector database. This type of database is specifically designed to efficiently store, manage, and search through large quantities of high-dimensional vector data.

\subsection{RAG-Based CTI Retrieval}

We proposed a new RAG-based search method tailored for gathering CTI from diverse sources. This method combines both standard search and NLP-based search techniques. 

The standard search, depicted in part 2 of Figure \ref{fig:RAG-Retrieval-component}, involves querying private databases commonly used by SOC teams. To automate this process, we first employed an IOC Parser tool to extract various IOCs, such as domain names, hashes, IP addresses, and URLs. For each IOC the system leverages APIs from private threat intelligence databases (VirusTotal in our case) to gather detailed contextual information, including historical malicious activity, geolocation, and reputation scores. This enriched data is then appended to the standard search context. For example, if an IP address is flagged, the system queries APIs to determine its association with botnets, historical attack patterns, and blacklist status.

The natural language processing (NLP)-based search, presented in part 3 of Figure \ref{fig:RAG-Retrieval-component}, retrieves the most relevant context from CTI reports by conducting searches in a knowledge base built from these. To retrieve the CTI reports most relevant to the detected incident in the alert, we propose a solution to enhance the alert. Raw alert data often lacks the contextual richness necessary for effective correlation. To address this, we leverage an LLM guided by a tailored prompt to generate an augmented alert. This prompt is designed to incorporate findings from an initial standard search—such as those conducted on VirusTotal during our experiments—into the alert. This process facilitates similarity-based searches by reformatting the alert into a structured and contextually relevant format.

Performing a search in the vector database involves calculating the similarity (e.g. cosine similarity) between the embeddings of the question (the Siem Alert) and document chunks (CTI).
To evaluate the similarity between texts, two key aspects must be defined: the method used to measure the similarity between embeddings, and the algorithm used to transform the text into embeddings, which represent the text in a vector space.
The system primarily employs Cosine Similarity, as defined in Equation \ref{eq:cosine_similarity}, to measure the similarity between embeddings. This method remains one of the most widely used techniques for assessing vector similarity. A score closer to 1 indicates a higher degree of similarity between embeddings.
To generate embeddings, we utilize Transformer-based models \cite{vaswani2017attention}, as they represent a significant breakthrough in NLP and consistently outperform previous approaches.

\begin{equation}
\text{similarity}(A, B) = \frac{\mathbf{A} \cdot \mathbf{B}}{\|\mathbf{A}\| \times \|\mathbf{B}\|} = \frac{\sum_{i=1}^n A_i \times B_i}{\sqrt{\sum_{i=1}^n A_i^2} \times \sqrt{\sum_{i=1}^n B_i^2}}
\label{eq:cosine_similarity}
\end{equation}


\subsection{Augmented Generation}
Following the retrieval step, a final prompt is designed for the LLM to generate an incident response strategy. This prompt incorporates both the SIEM alert and the retrieved data, which is a combination of standard search data (e.g. virusTotal results) and insights from the NLP search (CTI chunks). To achieve optimal results, we propose an adapted prompt that clearly defines the context of the alert, provides the necessary background from the retrieved data, and outlines the specific task. This ensures that the LLM produces a comprehensive and actionable incident response strategy tailored to the detected threat.
The final prompt is fed into a pre-trained LLM, as illustrated in Figure \ref{fig:RAG-Augmented-generation-component}, to generate the incident response plan text. In our solution, we leverage GPT LLM (Generative Pre-trained Transformer), which utilizes auto-regressive \cite{brown2020language} modeling to produce coherent and contextually relevant text. These models called decoder-only, correspond to the decoder part of the Transformer model, are ideal for text generation.


\section{EXPERIMENTS AND EVALUATION RESULTS} \label{sec:exp}


\subsection{System Setup}
We tested the proposed architecture on 100 alerts generated by the our company's SIEM system (LogPoint SIEM \cite{noauthor_logpoint_nodate}), randomly selected between August 10 and September 8, 2024. These real-world alerts served as inputs to our model, which we enriched with CTI to propose a coherent response for each incident.

\begin{figure}[hbt!]
  \centering
  \includegraphics[width=8cm]{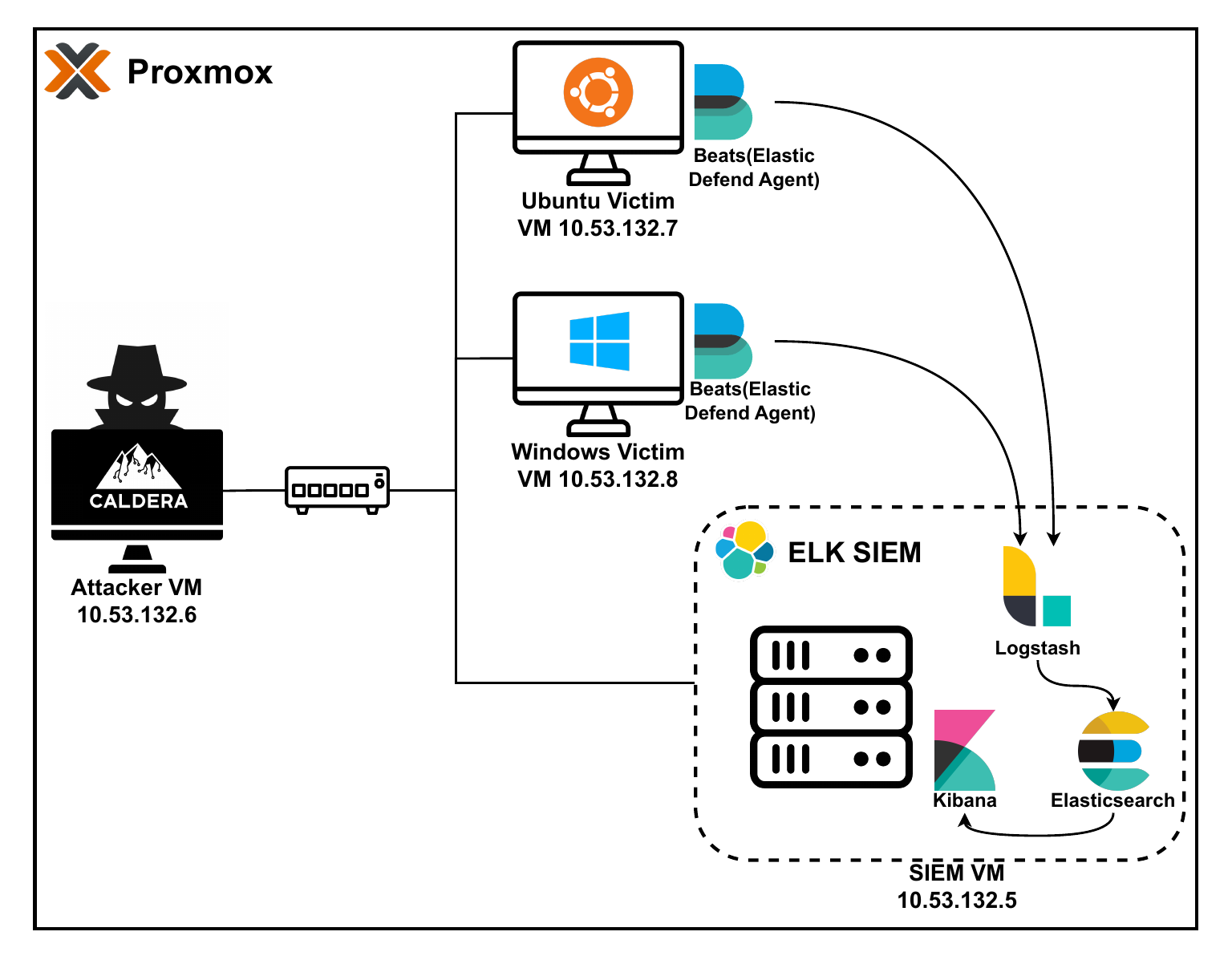}
  \caption{Test environment for alerts generation}
  \label{fig:test-env}
\end{figure}

In addition to real-world alerts, we manually generated alerts by emulating incidents within a controlled and isolated test environment, illustrated in Figure \ref{fig:test-env}. This environment was built on an internal Proxmox hypervisor, which hosted and managed the virtual machines (VMs) necessary for simulating attacks. We deployed the ELK stack \cite{noauthor_elk_nodate} as a dedicated SIEM solution on a separate virtual machine, configured within the same network as the target machines.
For attack simulation, we used a Linux virtual machine as the attacker, where we installed Caldera \cite{noauthor_caldera_nodate}, an attack simulation tool developed by MITRE. Caldera automates cyberattacks by replicating tactics and techniques (TTPs) from the MITRE ATT\&CK framework \cite{noauthor_mitre_nodate}. To ensure diverse log data, we deployed two types of target operating systems—Windows and Linux. Both target machines had Elastic Defend agents \cite{noauthor_elastic_nodate} installed to capture all security events.
Using Caldera, we emulated various advanced persistent threat (APT) attacks, illustrated in Table \ref{tab:sim-attacks}. These simulated attacks included a range of TTPs from the MITRE ATT\&CK Framework, targeting both Linux and Windows hosts. For instance, the Advanced Thief adversary employed techniques such as Automated Collection (T1119) and Exfiltration Over C2 Channel (T1041) on Linux machines to collect and exfiltrate sensitive data. Similarly, the Stowaway adversary utilized Process Discovery (T1057) and Process Injection (T1055.002) to hide its presence and evade detection on Windows systems. Detection rules were created within the SIEM to intercept these events and generate alerts. In total, we generated 10 simulated alerts, which were tested and evaluated using our solution. These alerts, combined with the 100 real-world alerts, provided a comprehensive dataset to thoroughly study and refine our proposed solution and analyze the results.

\begin{table}[h!]
\centering
\caption{Simulated Cyber Attacks in a Controlled Test Environment for Alert Generation} 
\resizebox{0.5\textwidth}{!}{
\begin{tabular}{|p{1.5cm}|p{3.5cm}|p{1.3cm}|p{5cm}|p{1cm}|}
\hline 
\textbf{Adversarie} & \textbf{Ability Name} & \textbf{Tactic} & \textbf{Technique} & \textbf{Host}\\ 
\hline 
Advanced Thief & Advanced File Search and Stager & Collection & T1119 Automated Collection & Linux \\ \hline
Advanced Thief & Compress staged directory & Collection & T1560.001 Archive Collected Data: Archive via Utility & Linux \\ \hline
Advanced Thief & Exfil staged directory & Exfiltration & T1041 Exfiltration Over C2 Channel & Linux \\ \hline
Stowaway & Discover injectable process & Discovery & T1057 Process Discovery & Windows \\ \hline
Stowaway & Inject Sandcat into process & Defense-Evasion & T1055.002 Process Injection: Portable Executable Injection & Windows \\ \hline
atomic & NMAP scan & Technical-Information-Gathering & T1254 Conduct active scanning & Linux \\ \hline
atomic & Access /etc/shadow (Local) & Credential-Access & T1003.008 OS Credential Dumping: /etc/passwd, /etc/master.passwd and /etc/shadow & Linux \\ \hline
Windows Worm \#1 & Collect ARP details & Discovery & T1018 Remote System Discovery & Windows \\ \hline
Super Spy & Find files & Collection & T1005 Data from Local System & Windows \\ \hline
Super Spy & Exfil staged directory windows & Exfiltration & T1041 Exfiltration Over C2 Channel & Windows \\ \hline
\end{tabular}}
\label{tab:sim-attacks}
\end{table}

For the standard search involving querying private databases, we used the VirusTotal database \cite{noauthor_virusTotal_nodate} via API, which allows searching for domain names, hashes, IP addresses, and URLs.
For the NLP-based search and vector database creation, we utilized public CTI reports from the APT repository on VX Underground \cite{noauthor_vx_nodate}, which contains papers and blogs (sorted by year) related to malicious campaigns, activity, or software associated with vendor-defined APT groups and toolsets. We retrieved these reports and performed text extraction from each CTI report in PDF format using PyMuPDF \cite{noauthor_pymupdf_nodate}, a high-performance Python library for extracting, analyzing, converting, and manipulating PDFs and other document formats. The extracted text was then chunked and embedded into a Chroma vector database.
For augmented generation, we used the GPT-4o model, as it is considered the best model according to the leaderboard platform on Hugging Face for text generation. Moreover, it accepts a large context length, which is crucial for our use case.


\subsection{Evaluation Method and Metrics}
To evaluate the RAG model, we opted for specific metrics commonly used for the assessment of Retrieval-Augmented Generation systems. These metrics are designed to measure various aspects of the model’s performance, focusing on how well it retrieves and generates information:
\begin{itemize}
    \item \textbf{Answer Relevance:} Is the response directly relevant to the query?
    \item \textbf{Context Relevance:} Is the retrieved context relevant and appropriately aligned with the query?
    \item \textbf{Groundedness:} Is the response supported by the retrieved context?

\end{itemize}
We propose using an automated evaluation method powered by LLMs to assess each metric for every alert. The idea is to employ other LLMs to evaluate each alert on a scale of 1 to 5 for each criterion. To accomplish this, we crafted specific prompts tailored to each metric. These prompts instruct the LLM to provide a score for each metric and simultaneously offer an explanation of why that score was given. To ensure the validity of our approach, we complemented the automated scoring system with manual evaluations performed by security experts, thereby validating the reliability and accuracy of the automated assessments.

\subsection{Experimental Results}
The evaluation consists of two parts. First, we assess the automated analysis of real alerts from the enterprise SIEM, evaluating the response quality. Then, we validate these results through both automated and manual evaluation on 10 simulated alerts generated in a controlled environment. This setup, with known incident causes, helps confirm the reliability of our scoring system. Finally, we compare results across both datasets.

\subsubsection{\textbf{Results of the real-world Alerts}} 
We performed the automatic evaluation on 100 real-world alerts reported by our security teams, using three open-source models. The models used include Mistral-large-2407 a larger variant of the Mistral model,  Llama-3.1-70B-Instruct, a larger variant of the LLaMA model from Meta, and a smaller LLaMA model Llama-3.2-3B-Instruct.

Table \ref{tab:llm_eval} displays the percentage ratings assigned by each model for each alert processed, along with the mean and variance. We observe that the models provide close mean scores, however their sensitivity varies. This variation is expected, as we included both large and small versions of different models.

Overall, our Retrieval-Augmented Generation (RAG) system demonstrates strong performance in answer relevance, achieving an average score close to 5 across all LLMs, and in groundedness (accuracy of responses), with an average score exceeding 4. These positive results are consistently reflected in the high ratings from all LLMs. However, context relevance scores are relatively low. The reduced score can likely be attributed to the inability to identify relevant context for certain alerts in either VirusTotal or CTI reports. This can be explained by the inclusion of false-positive alerts in the dataset, as well as the fact that, for some alerts, the IOCs are not recognized within the VirusTotal database. Furthermore, CTI reports often lack detailed information on specific campaigns. In future versions, we aim to extend this solution by incorporating additional trusted private databases, such as CrowdStrike, to enrich the standard search, as well as integrating more comprehensive CTI reports to improve overall context coverage.

To further investigate, we separated context relevance evaluations based on CTI reports and VirusTotal data. We observed similar, lower-than-global context relevance scores for each, indicating that both sources are complementary and collectively improve the final context relevance.

\begin{table}[h]
\centering
\caption{Automatic Evaluation of LLMs on Real-World Alerts}
\resizebox{0.5\textwidth}{!}{\begin{tabular}{lccccccc}
\hline
\textbf{Model \& Metric} & \textbf{1 (\%)} & \textbf{2 (\%)} & \textbf{3 (\%)} & \textbf{4 (\%)} & \textbf{5 (\%)} & \textbf{Mean} & \textbf{Variance} \\
\hline
Mistral-large Answer Relevance & 0.00 & 0.00 & 0.00 & 2.12 & 97.87 & \textbf{4.97} & 0.02 \\
Mistral-large Context Relevance [VT + CTI] & 0.00 & 20.00 & 51.11 & 22.22 & 6.66 & 3.15 & 0.66 \\
Mistral-large Context Relevance [VT only] & 17.77 & 3.33 & 44.44 & 27.77 & 6.66 & \textbf{3.02} & 1.28 \\
Mistral-large Context Relevance [CTI only] & 1.20 & 63.85 & 21.68 & 6.02 & 7.22 & 2.54 & 0.82\\
Mistral-large Groundedness & 0.00 & 0.00 & 2.46 & 41.97 & 55.55 & \textbf{4.53} & 0.29 \\
\hline
Llama-3.2-3B Answer Relevance & 0.00 & 1.00 & 23.00 & 25.00 & 51.0 & 4.26 & 0.71 \\
Llama-3.2-3B Context Relevance [VT + CTI] & 0.00 & 30.00 & 28.99 & 17.00 & 24.00 & 3.35 & 1.30 \\
Llama-3.2-3B Context Relevance [VT only] & 0.00 & 33.00 & 51.00 & 16.00 & 0.00 & 2.83 & 0.46 \\
Llama-3.2-3B Context Relevance [CTI only] & 0.00 & 20.00 & 31.00 & 28.99 & 20.00 & \textbf{3.49} & 1.04 \\
Llama-3.2-3B Groundedness & 0.00 & 3.00 & 28.00 & 16.00 & 53.00 & 4.19 & 0.89 \\
\hline
Llama-3.1-70B Answer Relevance & 0.00 & 0.00 & 0.00 & 5.00 & 95.0 & 4.95 & 0.04 \\
Llama-3.1-70B Context Relevance [VT + CTI] & 6.06 & 23.23 & 13.13 & 17.17 & 40.40 & \textbf{3.62} & 1.87 \\
Llama-3.1-70B Context Relevance [VT only] & 24.00 & 49.00 & 22.00 & 1.00 & 4.00 & 2.12 & 0.84 \\
Llama-3.1-70B Context Relevance [CTI only] & 2.00 & 20.00 & 64.00 & 1.00 & 13.00 & 3.03 & 0.80 \\
Llama-3.1-70B Groundedness & 0.00 & 0.00 & 18.00 & 13.00 & 69.00 & 4.51 & 0.60 \\
\hline
\end{tabular}}
\label{tab:llm_eval}
\end{table}


\subsubsection{\textbf{Results of the Simulated Alerts}}
In this section, we aim to validate the automated evaluation by incorporating a manual assessment conducted by cybersecurity experts.
This approach involves analyzing alerts generated through controlled simulations, where the incidents triggering each alert and their corresponding attacks are known. The objective is to assess the system in detail and validate the automated evaluation, which we intend to use later as a filter to retain only accurate responses.

We began with the automated evaluation, illustrated in Table \ref{tab:llm_eval_10}, which demonstrates results similar to the evaluation of 100 alerts. It shows strong performance in answer relevance, achieving an average score close to 5 across all LLMs, as well as groundedness (accuracy of responses), with an average score exceeding 4. Additionally, there is an improvement in context relevance compared to previous results, as all the alerts in this case are true positives. A detailed comparison between the two datasets will be explored in subsequent sections.

\begin{table}[h]
\centering
\caption{Automatic Evaluation of LLMs on Simulated Alerts}
\resizebox{0.5\textwidth}{!}{\begin{tabular}{lccccccc}
\hline
\textbf{Model \& Metric} & \textbf{1 (\%)} & \textbf{2 (\%)} & \textbf{3 (\%)} & \textbf{4 (\%)} & \textbf{5 (\%)} & \textbf{Mean} & \textbf{Variance} \\
\hline
Mistral-large-2407 Answer Relevance & 0.00 & 0.00 & 0.00 & 0.00 & 100 & \textbf{5.00} & 0.00 \\
Mistral-large-2407 Context Relevance [VT + CTI] & 0.00 & 0.00 & 22.22 & 0.00 & 77.77 & \textbf{4.55} & 0.69 \\
Mistral-large-2407 Context Relevance [VT only] & 22.22 & 11.11 & 22.22 & 11.11 & 33.33 & 3.22 & 2.39 \\
Mistral-large-2407 Context Relevance [CTI only] & 0.00 & 0.00 & 0.00 & 37.50 & 62.50 & \textbf{4.62} & 0.23 \\
Mistral-large-2407 Groundedness & 0.00 & 0.00 & 0.00 & 44.44 & 55.55 & 4.55 & 0.24 \\
\hline
Llama-3.2-3B-Instruct Answer Relevance & 0.00 & 0.00 & 0.00 & 20.00 & 80.00 & 4.80 & 0.15 \\
Llama-3.2-3B-Instruct Context Relevance [VT + CTI] & 0.00 & 10.00 & 20.00 & 30.00 & 40.00 & 4.00 & 1.00 \\
Llama-3.2-3B-Instruct Context Relevance [VT only] & 0.00 & 20.00 & 30.00 & 50.00 & 0.00 & \textbf{3.30} & 0.61 \\
Llama-3.2-3B-Instruct Context Relevance [CTI only] & 0.00 & 10.00 & 10.00 & 20.00 & 60.00 & 4.30 & 1.01 \\
Llama-3.2-3B-Instruct Groundedness & 0.00 & 0.00 & 10.00 & 10.00 & 80.00 & 4.70 & 0.41 \\
\hline
Llama-3.1-70B-Instruct Answer Relevance & 0.00 & 0.00 & 0.00 & 20.00 & 80.00 & 4.80 & 0.15 \\
Llama-3.1-70B-Instruct Context Relevance [VT + CTI] & 10.00 &  10.00 & 10.00 & 10.00 & 60.00 & 4.00 & 2.00 \\
Llama-3.1-70B-Instruct Context Relevance [VT only] & 40.00 & 20.00 & 10.00 & 10.00 & 20.00 & 2.50 & 2.45 \\
Llama-3.1-70B-Instruct Context Relevance [CTI only] & 0.00 & 10.00 & 20.00 & 0.00 & 70.00 & 4.30 & 1.21 \\
Llama-3.1-70B-Instruct Groundedness & 0.00 & 0.00 & 0.00 & 20.00 & 80.00 & \textbf{4.80} & 0.15 \\
\hline
\end{tabular}}
\label{tab:llm_eval_10}
\end{table}

For the expert evaluation, we submitted 10 alerts to a cybersecurity expert within our organization, as summarized in Table \ref{tab:manual_eval_10}. The results demonstrate strong performance across all metrics, with an average score exceeding 4. Compared to the automatic evaluation, the expert evaluation yields similar average scores for context relevance. However, the scores for answer relevance and groundedness are slightly lower, while still remaining near 4. 
These findings validate the proposed automatic evaluation methodology and further confirm the effectiveness of our approach from an operational perspective, as evaluated by an incident responder.


\begin{table}[h]
\centering
\caption{Expert Evaluation on Simulated Alerts}
\resizebox{0.5\textwidth}{!}{\begin{tabular}{lccccccc}
\hline
\textbf{Metric} & \textbf{1 (\%)} & \textbf{2 (\%)} & \textbf{3 (\%)} & \textbf{4 (\%)} & \textbf{5 (\%)} & \textbf{Mean} & \textbf{Variance} \\
\hline
Expert Answer Relevance & 0.00 & 0.00 & 0.00 & 80.00 & 20.00 & \textbf{4.20} & 0.17 \\
Expert Context Relevance & 0.00 & 0.00 & 20.00 & 50.00 & 30.00 & \textbf{4.10} & 0.54 \\
Expert Groundedness & 0.00 & 0.00 & 20.00 & 50.00 & 30.00 & \textbf{4.10} & 0.54 \\
\hline
\end{tabular}}
\label{tab:manual_eval_10}
\end{table}

\subsubsection{\textbf{Comparison of Results: Real-World Alerts vs. Simulated Alerts}}

Overall, the results show almost identical performance in Answer Relevance and Groundedness, both scoring very high and close to 5, with slightly better results for the simulated alerts. However, for Context Relevance, a significant improvement can be observed, increasing from a score of 3 to 4. This improvement is attributed to the presence of false positives in the real-world alerts, where the system fails to retrieve valid CTI. This limitation arises either during the standard search with VirusTotal, which fails to recognize any IOCs, or when CTI reports do not match any described campaigns. Consequently, the system generates a coherent result that is not based on the extracted context, leading to lower Context Relevance scores.

\section{CONCLUSION}\label{sec:con}
This paper introduces a novel intelligent incident response solution powered by large language models. Our solution effectively leverages Cyber Threat Intelligence through an innovative Retrieval-Augmented Generation architecture that integrates dual search techniques to contextualize and enrich incident data. The model performs NLP-based similarity searches within a CTI vector database, retrieving relevant documents or text segments to correlate incidents with historical cases. Additionally, it conducts standard searches via CTI APIs such as VirusTotal or CrowdStrike to access standardized contexts, facilitating correlation with data from private databases.
The proposed solution has been rigorously validated using a comprehensive evaluation framework that combines automated assessments using LLMs, and expert cross-validation by cybersecurity professionals. The evaluation demonstrates strong performance across critical metrics, including answer relevance, context relevance, and groundedness, using both real-world and simulated SIEM alerts. These results underscore the robustness and effectiveness of our proposed system.

In future research, we plan to extend our solution to a wider range of cybersecurity roles by tailoring outputs to their workflows and enhancing the model’s reasoning. Furthermore, we will address a critical dimension of the system's development: ensuring the security and resilience of the solution against adversarial attacks. By prioritizing these advancements, we aim to enhance the system's usability, adaptability, and trustworthiness in real-world applications.


\bibliographystyle{IEEEtran}
\bibliography{IEEEabrv,Bibliography}

\section{APPENDIX}


\onecolumn

\begin{tcolorbox}[title=Expansion Prompt Template]

You are a helpful cybersecurity expert.

Your task is to expand the given SIEM alert with additional context from VirusTotal to formulate a complete incident.

This expansion will facilitate similarity-based searches in Cyber Threat Intelligence (CTI) reports.  
You should explain the incident and detail the Indicators of Compromise (IoCs).

\textbf{SIEM Alert (Query):}  
\texttt{\{alert\}}

\textbf{VirusTotal Context:}  
\texttt{\{virustotal\_context\}}

\textbf{Answer:}  

\#\#\# Incident Overview

Based on the SIEM alert and VirusTotal context, this incident appears to involve incident description. The alert details suggest malicious activity that could be part of a larger campaign targeting a specific sector or industry. The VirusTotal analysis adds further clarity to the threat by identifying key indicators and behavioral patterns.

\#\#\# Indicators of Compromise (IoCs)

1. **Network Indicators**

   - Source IP: source\_ip (Possible attacker)  

   - Destination IP: destination\_ip (Potential target or intermediate server)  
   
   - Domain: domain (Linked to malicious activity)

2. **File Hashes** 

   - MD5:  
   
   - SHA1:  
   
   - SHA256:  
   
   - File associated with malware family: malware\_family (if identified)

3. **Behavioral Observations** 

   - behavior\_1  

   - behavior\_2

4. **VirusTotal Context** 

   - Detection Count: positives/total  
   
   - Associated Tags: tags  
   
   - Summary: analysis\_summary

\#\#\# Threat Hypothesis

This activity aligns with threat actor or group campaigns, which commonly use specific techniques or tactics. The observed indicators suggest potential motives or impact, and the behavior is consistent with related malware or known attack patterns.

\label{tab:expansion_template}
\end{tcolorbox}

\begin{tcolorbox}[title=Incident Response Prompt Template]
You are an Incident Responder (IR). 
Your Task is to provide a concise and relevant incident response strategy for the siem alert detected based on the context.\\

1- First, enrich and correlate the alert with VirusTotal results and cyber threat intelligence (CTI) context. \\

2- Then, Generate a detailed alert explanation when a match is found in VirusTotal or a Cyber Threat Intelligence (CTI) document. Include the full name of the matched document or report, and provide a comprehensive explanation of the potential attack, including its possible purpose, method of operation, and implications for the targeted system or organization. \\

3- Finaly, propose a clear and actionable incident response strategy tailored to the specific incident.\\

Your response should be clear, concise, and focused on the incident. If the answer cannot be deduced from the context, do not give an answer.\\

Incident (SIEM LOG): \{question\}\\

virustotal Results : \{virustotal\_context\}\\

CTI documents : \{context\}\\

Output::
\label{tab:IR_templates}
\end{tcolorbox}

\begin{tcolorbox}[title=Answer Relevance Prompt Template]
Task Description:
You will evaluate how well the generated response directly addresses the SIEM alert.\\

Instructions:\\
Assess the relevance of the response to the SIEM alert based on the following:

    Does the response focus on the key aspects of the alert?\\
    Is the response aligned with the nature of the alert (e.g., malware, phishing, intrusion)?\\
    Does it provide actionable insights or explanations that match the alert's context?

Scoring (1 to 5):\\
    1: The response is not relevant at all.\\
    2: The response is somewhat relevant but does not directly address the SIEM alert.\\
    3: The response is moderately relevant; it addresses some aspects but lacks focus.\\
    4: The response is mostly relevant with minor gaps.\\
    5: The response is highly relevant and fully addresses the SIEM alert.

**You must provide the Total Rating.**

Answer:::\\
Evaluation: (Explain why the response is or isn’t relevant)\\
Total Rating: (Provide a rating from 1 to 5)\\

SIEM Alert (Query): \{alert\}\\
Generated Response: \{response\}\\

Output:::
\label{tab:eval_AR_templates}
\end{tcolorbox}

\begin{tcolorbox}[title=Context Relevance Prompt Template]
Task Description:
You will evaluate whether the response appropriately considers the broader security context based on available information.

(Is the context useful for enriching the SIEM alert?)\\

Instructions:\\
Assess the context relevance of the response based on the following:\\
    Does the response consider the larger security implications of the alert?\\
    Is the explanation aligned with real-world attack techniques and threat intelligence?\\
    Does it make reasonable connections between the alert, VirusTotal results, and CTI data?

Scoring (1 to 5):\\
    1: The response lacks any meaningful context or is misleading.\\
    2: The response includes some context but misses key connections.\\
    3: The response considers context but is not well-integrated with the provided data.\\
    4: The response is contextually relevant with only minor gaps.\\
    5: The response fully integrates and applies context appropriately.

**You must provide the Total Rating.**

Output:::\\
Evaluation: (Explain your reasoning for the context relevance rating)\\
Total Rating: (Provide your rating here, from 1 to 5)\\

SIEM Alert (Query): \{alert\}\\
Context: \{context\}\\

Answer:::
\label{tab:eval_CR_templates}
\end{tcolorbox}

\begin{tcolorbox}[title=Groundedness Prompt Template]

Task Description:
You will evaluate whether the response is properly supported by the given VirusTotal results and CTI documents.\\
(Is the response well-supported by the context?)\\

Instructions:\\
Assess the groundedness of the response based on the following:\\
    Does the response correctly use information from VirusTotal and CTI sources?\\
    Is there any unsupported or hallucinated information in the response?\\
    Does the response cite relevant CTI documents or VirusTotal results appropriately?

Scoring (1 to 5):

    1: The response contains hallucinated or unsupported information.\\
    2: The response includes some relevant information but introduces inaccuracies.\\
    3: The response is mostly based on sources but has minor unsupported claims.\\
    4: The response is well-grounded with only slight inconsistencies.\\
    5: The response is fully supported by VirusTotal and CTI documents.

**You must provide the Total Rating.**

Answer:::\\
Evaluation: (Explain your reasoning for the groundedness rating)\\
Total Rating: (Provide your rating here, from 1 to 5)\\

Incident Response Strategy (Response): \{response\}\\
Context: \{context\}\\

Output:::
\label{tab:eval_GND_templates}
\end{tcolorbox}

\twocolumn
\vspace{12pt}

\end{document}